\documentclass[pre,twocolumn,superscriptaddress,showpacs,floatfix,amssymb,amsmath]{revtex4}

\usepackage[pdftex]{graphicx}
\usepackage{dcolumn}
\usepackage{bm}
\usepackage{color}
\usepackage{amsmath}
\usepackage{graphicx}
\usepackage{amssymb}
\usepackage{mathrsfs}
\usepackage{bbm}
\usepackage{epsfig}

\newcommand{\be}{\begin{eqnarray}}
\newcommand{\ee}{\end{eqnarray}}

\begin{document}
%


\title{
Peierls-Nabarro Barrier and Protein Loop Propagation
}

\author{Adam K. Sieradzan}
\email{adasko@sun1.chem.univ.gda.pl}
\affiliation{Department of Physics and Astronomy, Uppsala University, {\AA}ngstr\"{o}mlaboratoriet, L\"{a}gerhyddsv\"{a}gen 
1, 751 20 Uppsala, Sweden}
\affiliation{Faculty of Chemistry, University of Gda\'nsk, Wita Stwosza 63,
80-952 Gda\'nsk, Poland}
\author{Antti Niemi}
\email{Antti.Niemi@physics.uu.se}
\affiliation{Department of Physics and Astronomy, Uppsala University, {\AA}ngstr\"{o}mlaboratoriet, L\"{a}gerhyddsv\"{a}gen 
1, 751 20 Uppsala, Sweden}
\affiliation{Laboratoire de Mathematiques et Physique Theorique CNRS UMR 6083, F\'ed\'eration Denis Poisson, Universit\'e 
de Tours, Parc de Grandmont, F37200 Tours, France}
\affiliation{Department of Physics, Beijing Institute of Technology, Haidian District, Beijing 100081, People‚Äôs Republic of 
China}
\author{Xubiao Peng}
\email{xubiaopeng@gmail.com}
\affiliation{Department of Physics and Astronomy, Uppsala University, {\AA}ngstr\"{o}mlaboratoriet, L\"{a}gerhyddsv\"{a}gen 
1, 751 20 Uppsala, Sweden}

\begin{abstract}
\noindent
When a self-localized quasiparticle excitation propagates along  
a discrete one dimensional lattice,  it becomes subject to a dissipation that 
converts the  kinetic energy 
into lattice vibrations. Eventually the kinetic energy does no longer enable the  
excitation to cross over the minimum energy 
barrier  between neighboring sites,  and the excitation becomes 
localized within a lattice cell.  
In the case of a protein, the  lattice structure consists of the C$^\alpha$ backbone.
The self-localized quasiparticle excitation is the elemental
building block of loops. It can be modeled by a kink which solves a 
variant of the discrete non-linear Schr\"odinger equation (DNLS). 
We study the propagation of such a kink in the case of protein G related 
albumin-binding domain,  using the UNRES coarse-grained molecular dynamics 
force field.   We estimate the height of the energy barriers 
the kink needs to cross over, in order to propagate  
along the backbone lattice. We analyse how these barriers gives rise to both 
stresses and reliefs which control the  kink movement. 
For this, we deform a natively folded protein structure 
by parallel translating the kink along the 
backbone away from its native position.  
We release the transposed kink, and we follow how it propagates along the backbone towards the
native location. We observe that the dissipative forces which are exerted 
on the kink by the various energy barriers,  have a  pivotal  role in determining how 
a protein folds towards its native state.
\end{abstract}


\maketitle

\section{Introduction}
The protein folding problem has been around for a long time \cite{anfinsen_1975}, with 
various incarnations  \cite{dill_2008,dill_2012,pettitt_2013}. 
Numerous proposals have been presented, to explain how the folding patterns emerge and advance. 
Examples include
nucleation-condensation \cite{lopez_1996}, hydrophobic collapse \cite{zhou_2004} 
and diffusion-collision controlled processes \cite{bashford_1988,karplus_1994}. 
More recently, a kink-based approach has been proposed \cite{chernodub_2010,molkenthin_2011}
and tested successfully  using a coarse-grained  force-field \cite{krokhotin_2012,krokhotin_2014}.
At the same time,  protein misfolding 
continues to be vigorously investigated \cite{allan_2002}.
A misfolded protein can lead  to serious complications,
a misfolding transition is now  widely accepted as the
cause for diseases like type-II diabetes, many forms of cancer, and dementia including Alzheimer's
 \cite{soto_1998}. 

In the present article we investigate the phenomenon of {\it barrier penetration} as an example of those
physical mechanisms that contribute to  protein folding and misfolding transitions. For this we inquire what happens when a 
protein loop fragment becomes displaced from its native position. How does it find its way back
to the native position? The understanding  of the dynamical processes involved should contribute
 to  our comprehension of  protein
folding, including  the reasons why and how a misfolding transition takes place.

As a concrete example we consider an albumin-binding domain which is related to the  protein G.
This is a fairly generic $\alpha$-helical protein, the native state involves three helices which are 
separated by two short loops; we use
the Protein Data Bank (PDB)  \cite{berman_2000}
entry code 1GAB. It consists of 53 amino acids, the first loop is located between sites 19-24 and 
the second loop is located between sites 33-37.
The reason for choosing 1GAB as our concrete example is, that the UNRES force field, on which our investigations are based,
has been optimised for this particular protein.

In our analysis of 1GAB we first follow \cite{chernodub_2010,molkenthin_2011}, to  conclude that each of
the two loops of 1GAB are modeled by  a single kink that solves 
a generalized discrete non-linear Schr\"odinger equation \cite{faddeev_2007,ablowitz_2004,kevrekidis_2009}.
We then parallel translate the second loop in the PDB structure by five residues towards the first loop,  
with no change in the remaining residues.  We release the loop so that it is free to move
along the backbone. This  instigates 
a process where the kink attempts to propagate back to  its original posture. Thus we have a prototype example of the
Peierls-Nabarro scenario where a coherent and self-localized quasiparticle excitation 
propagates along a discrete one dimensional lattice \cite{peierls_1940,nabarro_1947,nabarro_1997}. Now
realised in the case of the protein backbone.

\section {Methods}

Our analysis is based on the United Residue (UNRES) force field
\cite{liwo_1997,liwo_1997_02,liwo_2001,liwo_2007,liwo_2008,kozlowska_2010_02,sieradzan_2012a,sieradzan_2012c}.
It is freely available at 
\begin{equation}
{\tt http://www.unres.pl}
\label{unres}
\end{equation} 
UNRES  is a carefully grafted coarse-grained force field that 
can be utilized to simulate the time evolution of a displaced kink along the protein backbone, in a highly realistic 
manner. UNRES has a very strong predictive power \cite{he_2013}.  It has also been 
optimised to reproduce the correct thermodynamic properties, in a scenario like the one that will be studied here. 
Moreover, UNRES has a  proven track record of already successfully modelling  the kink formation in 
an $\alpha$-helical protein \cite{krokhotin_2014}, like the one studied here. 
With the presently available computer power, a coarse-grain force-field such as UNRES enables us to
produce numerous full-length trajectories to probe the dynamical details in terms of a statistical analysis, which we
needed  in particular when  determining the temperature dependence of the Peierls-Nabarro barrier.

\subsection{ \it Protein backbone geometry}
\label{sect:backbone}

\begin{figure}
  \begin{center}
    \resizebox{8cm}{!}{\includegraphics[]{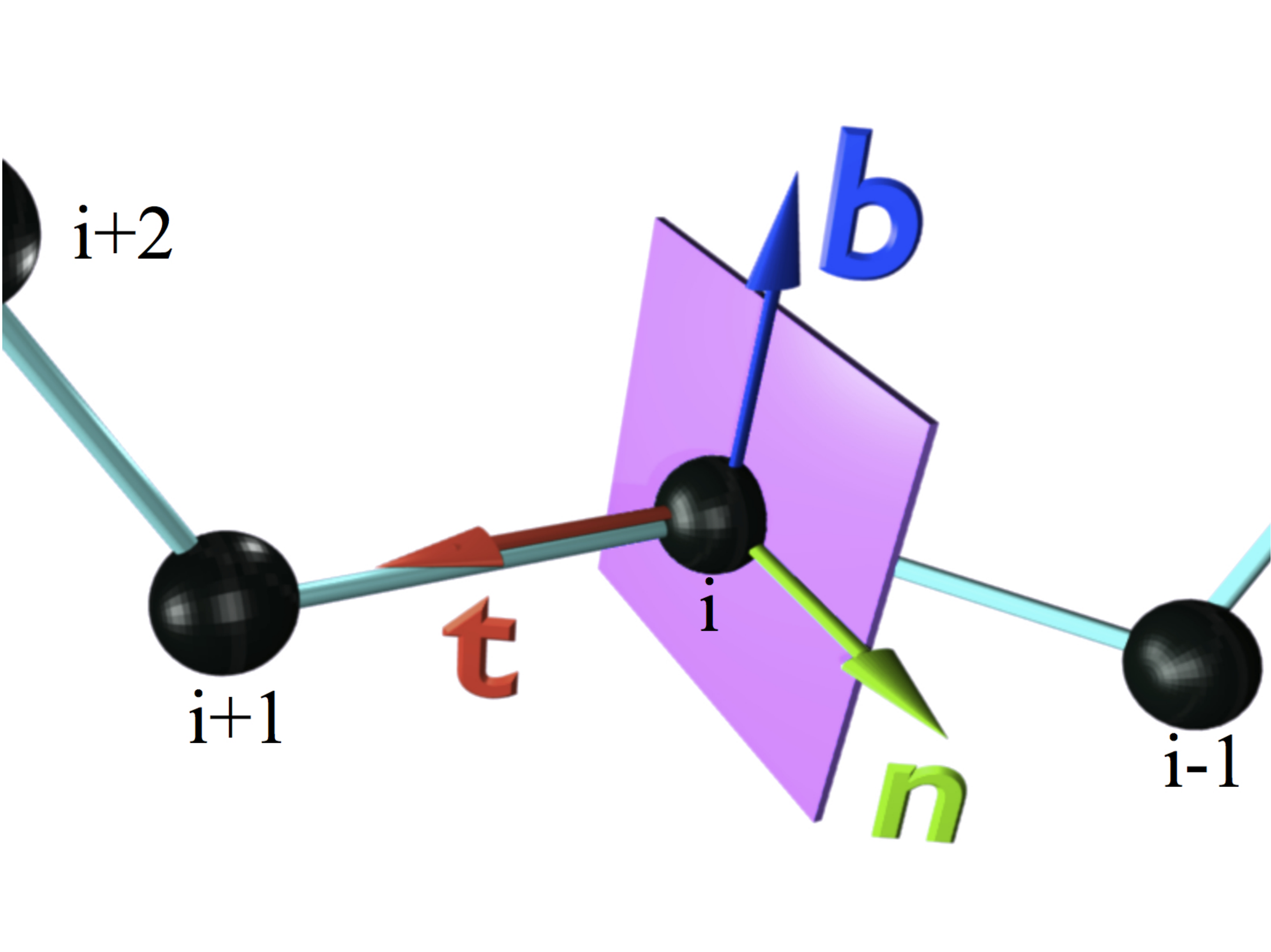}}
    \caption{(Color online) Frenet frames along the C$^\alpha$ backbone lattice.
   }
    \label{fig-1}
  \end{center}
\end{figure}

\begin{figure}
  \begin{center}
    \resizebox{8cm}{!}{\includegraphics[]{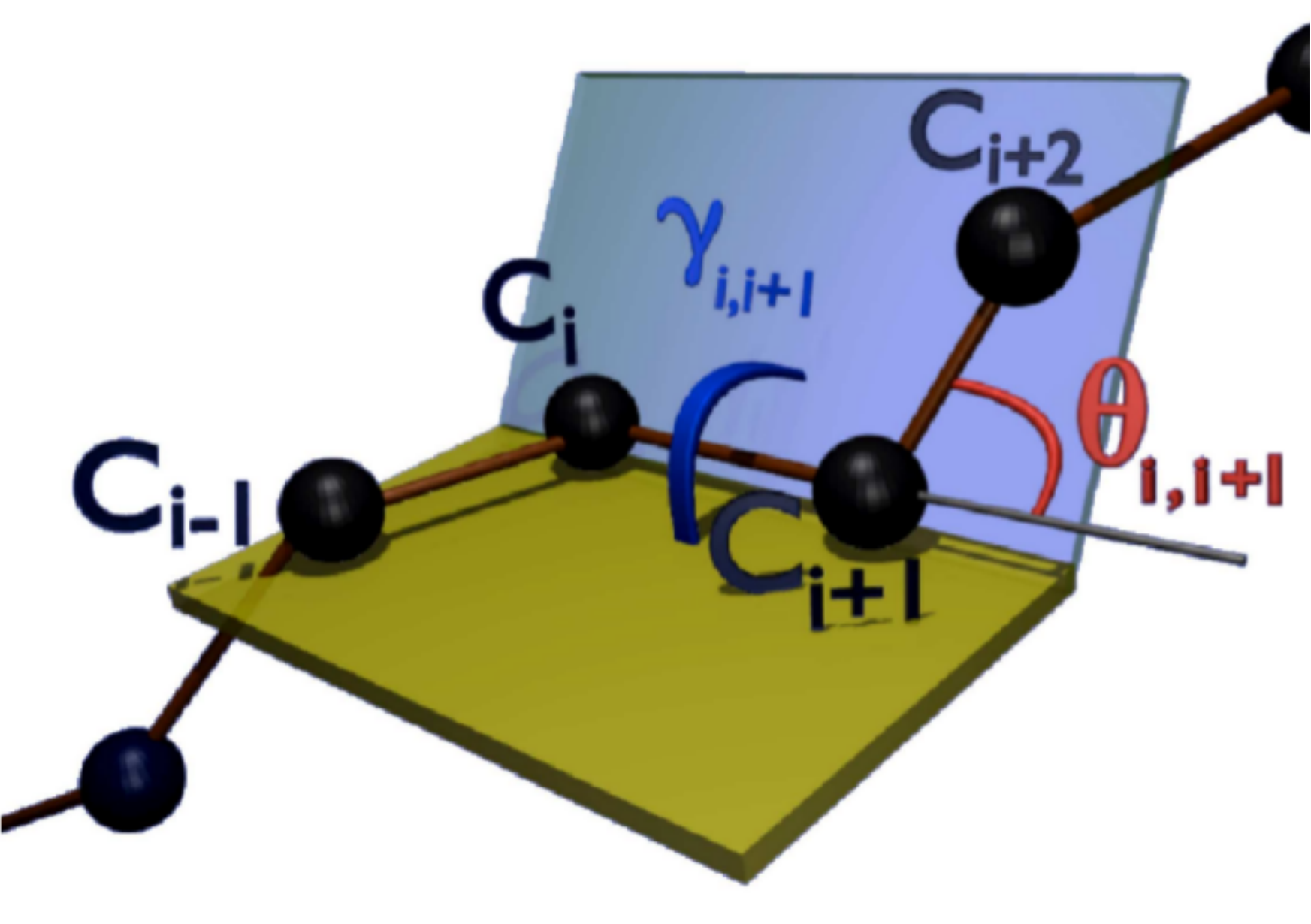}}
    \caption{(Color online) Definition of bond ($\theta_i$) and torsion ($\gamma_i$) angles, along the discrete C$^\alpha$ string.
   }
    \label{fig-2}
  \end{center}
\end{figure}

We describe  the skeletal  C$^\alpha$ backbone lattice, using
the geometrically determined Frenet frames  \cite{hu_2011-2}.
These frames depends {\it only} on the position 
of  the C$^\alpha$ carbons, with  coordinates ${\bf r}_i$ where $i=1,...,n$ labels the residues. 
At a given residue, the orthonormal Frenet frame consists of the backbone tangent ($\mathbf t$), 
binormal ($\mathbf b$), and normal ($\mathbf n$) vectors. These are defined by equations (\ref{eq:t}), 
(\ref{eq:b}), and (\ref{eq:n}), respectively, and shown in figure \ref{fig-1}.
\begin{equation}
\mathbf t_i = \frac{ {\bf r}_{i+1} - {\bf r}_i  }{ |  {\bf r}_{i+1} - {\bf r}_i | }
\label{eq:t}
\end{equation}
\begin{equation}
\mathbf b_i = \frac{ {\mathbf t}_{i-1} \times {\mathbf t}_i  }{  |  {\mathbf t}_{i-1} \times {\mathbf t}_i  | }
\label{eq:b}
\end{equation}
 \begin{equation}
\mathbf n_i = \mathbf b_i \times \mathbf t_i ~~~~
\label{eq:n}
\end{equation}
For {\it trans}-peptide groups the distance between the consecutive C$^\alpha$ atoms is 
close to the constant value
\[
| {\mathbf r}_{i+1} - {\mathbf r}_i | \ = \ \Delta \ \approx 3.8 {  \rm \AA}
\] 
Thus the backbone lattice 
becomes entirely described by virtual-bond-valence angles $\theta$ and virtual-bond-dihedral (torsion)
angles $\gamma$; see figure \ref{fig-2}.
\begin{eqnarray}
\theta_{i+1,i}  \equiv \ \theta_i   \! & = & \!  \arccos (  \mathbf t_{i+1} \cdot \mathbf t_i  ) \label{eq:bond} \\
\gamma_{i+1,i} \equiv  \gamma_{i}  \! & =  & \!  {\rm sgn}[(\mathbf b_{i}\times \mathbf b_{i+1})\cdot\mathbf t_i]  
 \arccos (\mathbf  b_{i+1} \! \cdot \mathbf b_i ) ~
\label{eq:torsion}
\end{eqnarray}
The frame vectors at site $\mathbf r_{i+1}$ are expressed in terms of the frame vectors at site $\mathbf r_i$ 
and the angles $\theta$ and $\gamma$ using a transfer matrix,
\begin{equation}
\left( \begin{matrix} {\bf n}_{i+1} \\  {\bf b }_{i+1} \\ {\bf t}_{i+1} \end{matrix} \right)
= 
\left( \begin{matrix} \cos\theta \cos \gamma & \cos\theta \sin\gamma & -\sin\theta \\
-\sin\gamma & \cos\gamma & 0 \\
\sin\theta \cos\gamma & \sin\theta \sin\gamma & \cos\theta \end{matrix}\right)_{\hskip -0.1cm i+1 , i}
\left( \begin{matrix} {\bf n}_{i} \\  {\bf b }_{i} \\ {\bf t}_{i} \end{matrix} \right) 
\label{eq:DFE2}
\end{equation}
The C$^\alpha$ lattice 
is constructed as follows,
\begin{equation}
\mathbf r_k = \sum_{i=0}^{k-1} |\mathbf r_{i+1} - \mathbf r_i | \cdot \mathbf t_i \ \equiv \ \Delta \sum_{i=0}^{k-1} \mathbf t_i
\label{eq:dffe}
\end{equation}
The normal and binormal vectors ($\mathbf n_i, \mathbf b_i$) are both absent
in equation (\ref{eq:dffe}), only the tangent vector $\mathbf t_i$ appears. Therefore, if these two vectors 
are simultaneously rotated around  the vector $\mathbf t_i$, they continue to constitute an equally
good framing.  In particular a frame rotation by $\pi$ determines the discrete $\mathbb Z_2$ 
transformation
\begin{equation}
\begin{matrix}
\ \ \ \ \ \ \ \ \ \gamma_{i }  & \to &  \hskip -2.5cm \gamma_{i} - \pi  \\
\ \ \ \ \ \ \ \ \ \theta_{k} & \to  &  - \ \theta_{k} \ \ \ \hskip 1.0cm  {\rm for \ \ all} \ \  k \geq i 
\end{matrix}
\label{eq:dsgau}
\end{equation}
that leaves (\ref{eq:dffe}) intact. This $\mathbb Z_2$ transformation has been 
used extensively to analyze proteins   
\cite{hu_2011-2,lundgren_2013,hu_2013,krokhotin_2013,krokhotin_2013-2,krokhotin_2012,krokhotin_2014}
and will also be used here.  

\subsection{\it Kink  and protein geometry} 
\label{sect:solitons}

The virtual
bond angles  (\ref{eq:bond}) and virtual torsion angles (\ref{eq:torsion}) of the C$^\alpha$ backbone
are mutually correlated, by the atomic level interactions 
along the protein chain. We may choose these angles as  
the order parameters, that characterise  
the thermodynamical  state of the protein in terms of its geometric composition.
In  \cite{niemi_2003,danielsson_2010,chernodub_2010,molkenthin_2011,krokhotin_2012} 
it has been shown that in the  limit of  slowly varying variables,
 the full all-atom thermodynamical free energy $F$ can be expanded in terms
of these order parameters, in the universal sense of  Kadanoff and Wilson 
\cite{kadanoff_1966,wilson_1971}, as follows:
\[
F  = - \sum\limits_{i=1}^{N-1}  2\, \theta_{i+1} \theta_{i}  + 
\]
\begin{equation}
+ \sum\limits_{i=1}^N
\biggl\{  2 \theta_{i}^2 + \lambda\, (\theta_{i}^2 - m^2)^2  
\
\  
+ \frac{q}{2} \, \theta_{i}^2 \gamma_{i}^2   
- p \,  \gamma_{i}   +  \frac{r}{2}  \gamma^2_{i} 
\biggr\} \ + \dots
\label{E1old}
\end{equation}
In equation (\ref{E1old}) $\lambda$, $q$, $p$, $r$, and $m$ 
depend on the atomic level physical properties and the chemical 
microstructure of the protein and its environment. In principle these parameters can 
be computed from this knowledge.

The kink that constitutes the elemental building block of a protein loop 
arises from (\ref{E1old}) as follows: 
The angles $\gamma_i$ are first determined as  functions of the 
angles $\theta_i$,
\begin{equation}
\gamma_{i} [\theta] \ =  \ \frac{u }{1 + v \, \theta^2_{i} } 
\label{tauk} 
\end{equation}
with $u=p/r$ and $v=q/r$. The equation (\ref{tauk}) is inserted into
equation (\ref{E1old}), to  eliminate the $\gamma_i$ in favor of the 
$\theta_i$. This gives a set of equations (\ref{nlse}) 
for the $\theta_i$ 
\begin{equation}
\theta_{i+1} = 2\theta_i - \theta_{i-1} + \frac{ d V [\theta]}{d\theta_i^2} \theta_i  \ \ \ \ \ (i=1,...,N)
\label{nlse}
\end{equation}
where $\theta_0 = \theta_{N+1}=0$
and 
\begin{equation}
V [\theta]  =  - \ \frac{u^2r}{2(1 + v \, \theta^2) } - 2\lambda m^2  \theta^2
+ \lambda \, \theta^4
\label{U}
\end{equation}
Here we recognize the structure of a generalized discrete non-linear Schr\"odinger (DNLS)  equation
\cite{faddeev_2007,ablowitz_2004,kevrekidis_2009,chernodub_2010,molkenthin_2011}
In the case of proteins, $u$ is small and $\lambda m^2>0$ so that there 
are two local minima in $V[\theta]$, with $\theta = \pm \theta_0 \approx \pm m$. 
The kink is a solution of (\ref{nlse}), (\ref{tauk}) that interpolates between the two minima.
The explicit form of the kink is not known to us, in terms of elementary functions.
But it can be constructed  numerically by following the iterative procedure of ref. 
\cite{molkenthin_2011}. Note that a ground state configuration where $\theta_i$ is constant with
\begin{equation}
\left\{ \ \begin{matrix} \theta_i  & \approx & \frac{\pi}{2} \\
\gamma_i & \approx & 1
\end{matrix} \right. \ \ \ \ \ \ \ {\rm  (radians)}
\label{alphahel}
\end{equation}
describes a right-handed $\alpha$-helix, while
\begin{equation}
\left\{ \ \begin{matrix} \theta_i  & \approx & 1 \\
\gamma_i & \approx & \pm \pi
\end{matrix} \right.  \ \ \ \ \ \ \ {\rm  (radians)}
\label{betas}
\end{equation}
describes a $\beta$-strand.  For a kink, the ($\theta_i,\gamma_i$) profile interpolates with these  
regular secondary structures. 

For a given protein structure, we train the energy function (\ref{E1old}) to describe the backbone geometry
in terms of a multi-kink solution of  (\ref{nlse}), (\ref{tauk}). For this we use the packages {\it Propro} and {\it GaugeIT},
described at 
\[
{\tt http://www.folding-protein.org}
\]
In the case of 1GAB, we use the averaged NMR structure.
\begin{figure}
  \begin{center}
    \resizebox{8cm}{!}{\includegraphics[]{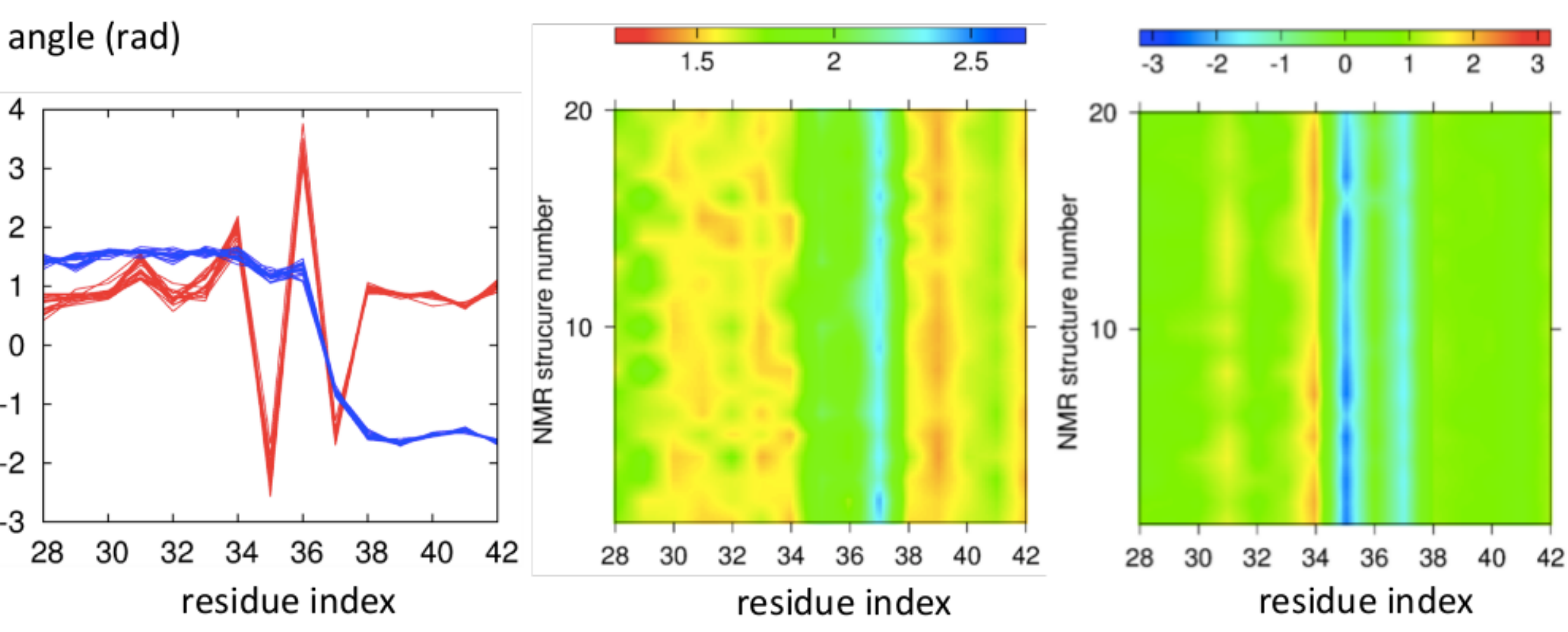}}
    \caption{(Color online) (Left) The $\theta$ (blue) and $\gamma$ (red) profile of 1GAB, average over all NMR structures, 
    after $\mathbb Z_2$ transformation. In (centre) and (right) we show the variations in $\theta$ and $\gamma$ profiles, 
    respectively, over the different NMR structures of 1GAB; colour intensity 
    determines the value of the angle in radians, according to
    the scale on top.
   }
    \label{fig-3}
  \end{center}
\end{figure}
 In figures \ref{fig-3} we describe the profile of the 
second loop along the
1GAB backbone, averaged over all NMR structures in PDB. 
In the figure  \ref{fig-3} (left)  we have implemented the $\mathbb Z_2$ 
transformation (\ref{eq:dsgau}).
It reveals that the loop is a single kink, that appears as a domain wall between two 
$\alpha$-helical structures.

\subsection{\it The Peierls-Nabarro  barrier}
\label{sect:methods:CG}

In a protein, a loop propagation instigates a structural deformation.
 In crystallographic protein structures the loops are located at definite positions along the backbone chain. 
Thus, in the case of a static, crystallographic protein the translation
 invariance of the kink along the C$^\alpha$ lattice is clearly broken.
The breakdown of translation invariance in the case of a lattice kink is in fact generic:  
Whenever a self-localized quasiparticle 
excitation such as a kink moves along the  lattice, waves are emitted  in its wake as vibrations in
the lattice structure.  These waves drain the kinetic energy of the quasiparticle  excitation, 
causing it to decelerate.  Eventually the kinetic energy becomes depleted to the extent that 
the excitation can no longer cross over the energy barriers between lattice cells, and instead it becomes
localizes to a particular lattice cell. 

Normally, a localized quasiparticle  excitation has only two stationary positions relative to the lattice structure.   
Moreover, in a simple model these energy extrema are often positioned symmetrically,
either at a lattice site or half-way between two neighboring 
lattice sites. One of the stationary positions is a local minimum of the potential energy, while 
the other is a  saddle point.  The energy difference between these two 
configurations determines the basal kinetic energy that 
the quasiparticle excitation needs, in order to translate itself from one lattice cell to another. 
This minimum energy difference 
is called the {\it Peierls-Nabarro barrier} (PNB) of the 
excitation \cite{peierls_1940,nabarro_1947,nabarro_1997}.

\subsection{\it Simulation details}

\begin{figure}
  \begin{center}
    \resizebox{8cm}{!}{\includegraphics[]{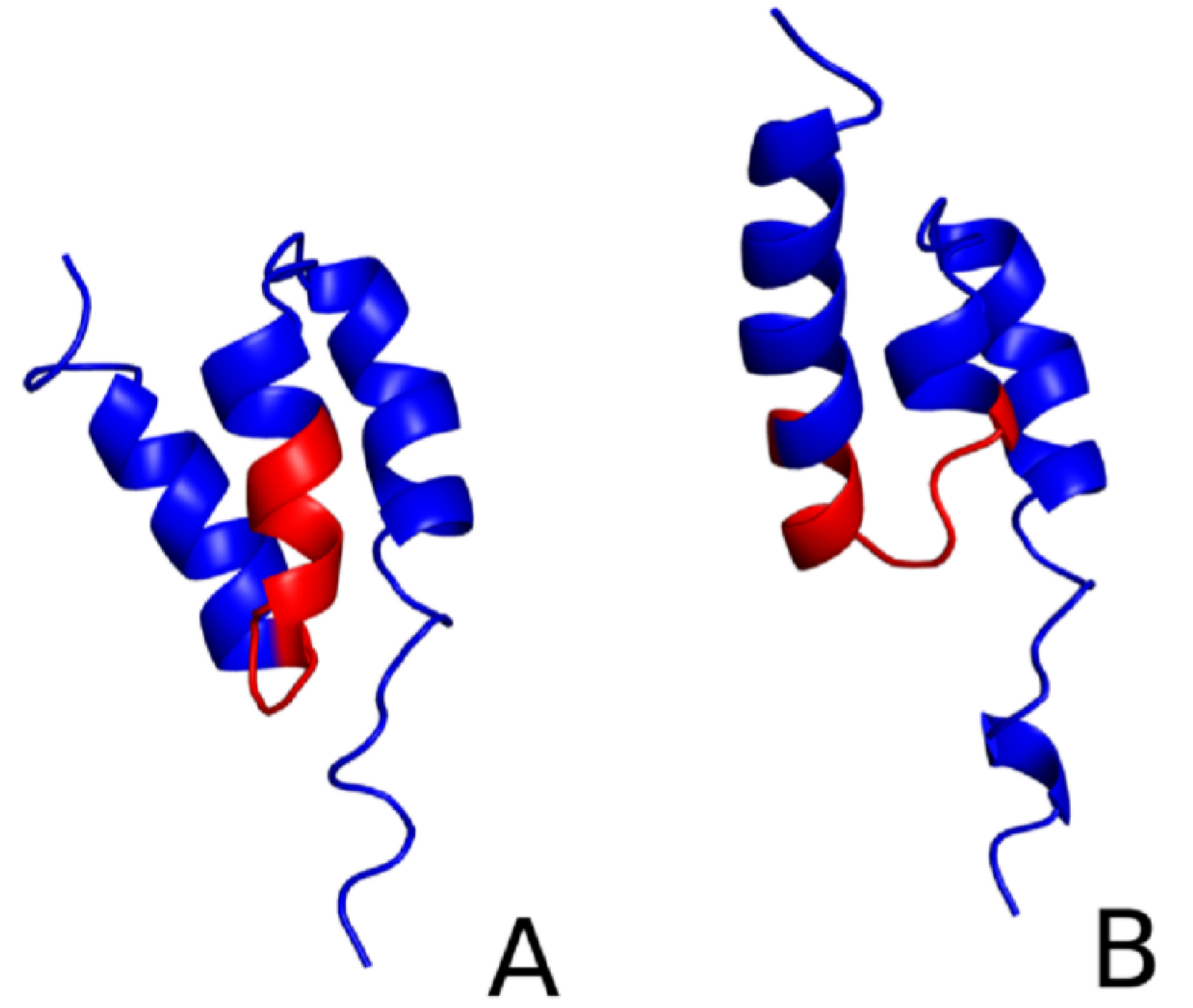}}
    \caption{(Color online) The protein G-related albumin-binding domain (PDB code: 1GAB) 
    native structure (A), structure with shifted loop by five residues (B). The red fragment 
    shows where the two structures differ.
   }
    \label{fig-4}
  \end{center}
\end{figure}

We use the UNRES force field (\ref{unres}) in combination with 
the Molecular Dynamics (MD) protocol described in \cite{khalili_2005,khalili_2005_02,liwo_2005}. 
The initial condition for our simulations is  the PDB structure 1GAB, where we have
parallel transported  the second loop 
fragment Ala$^{35}$-Val$^{38}$ 
to the position of  Gln$^{30}$-Gln$^{33}$. 
Figure  \ref{fig-4} 
compares the initial configuration with the crystallographic 1GAB structure.
Note that the ($\theta_i,\gamma_i$) spectrum remains intact during the parallel transport,  except for the uniform loop translation.  
The ensuing C$^\alpha$ configuration is converted into an all-atom structure using a combination of 
Pulchra \cite{rotkiewicz_2008} and SCWRL4 \cite{krivov_2009}. Our choice  
to translate the loop fragment  by five residues follows from an inspection of the all-atom
steric hindrances that are incurred
during the translation process.

We have simulated a total of 704  
canonical molecular dynamics trajectories, at 11 different temperature values   from 270K to 320K  
by steps of  5K.  Thus there are 64 trajectories, at each temperature value.  
Each  trajectory consists of 10 million iterations, with UNRES
time step value 4.89 fs. Since the very fast atomic level motions are averaged by the UNRES force field,  the corresponding {\it in vivo}  time is around $\sim$ 50 $\mu$s \cite{khalili_2005_02,liwo_2005,sieradzan_2012a}.

\subsection{\it Analysis details}

\begin{figure}
  \begin{center}
    \resizebox{8cm}{!}{\includegraphics[]{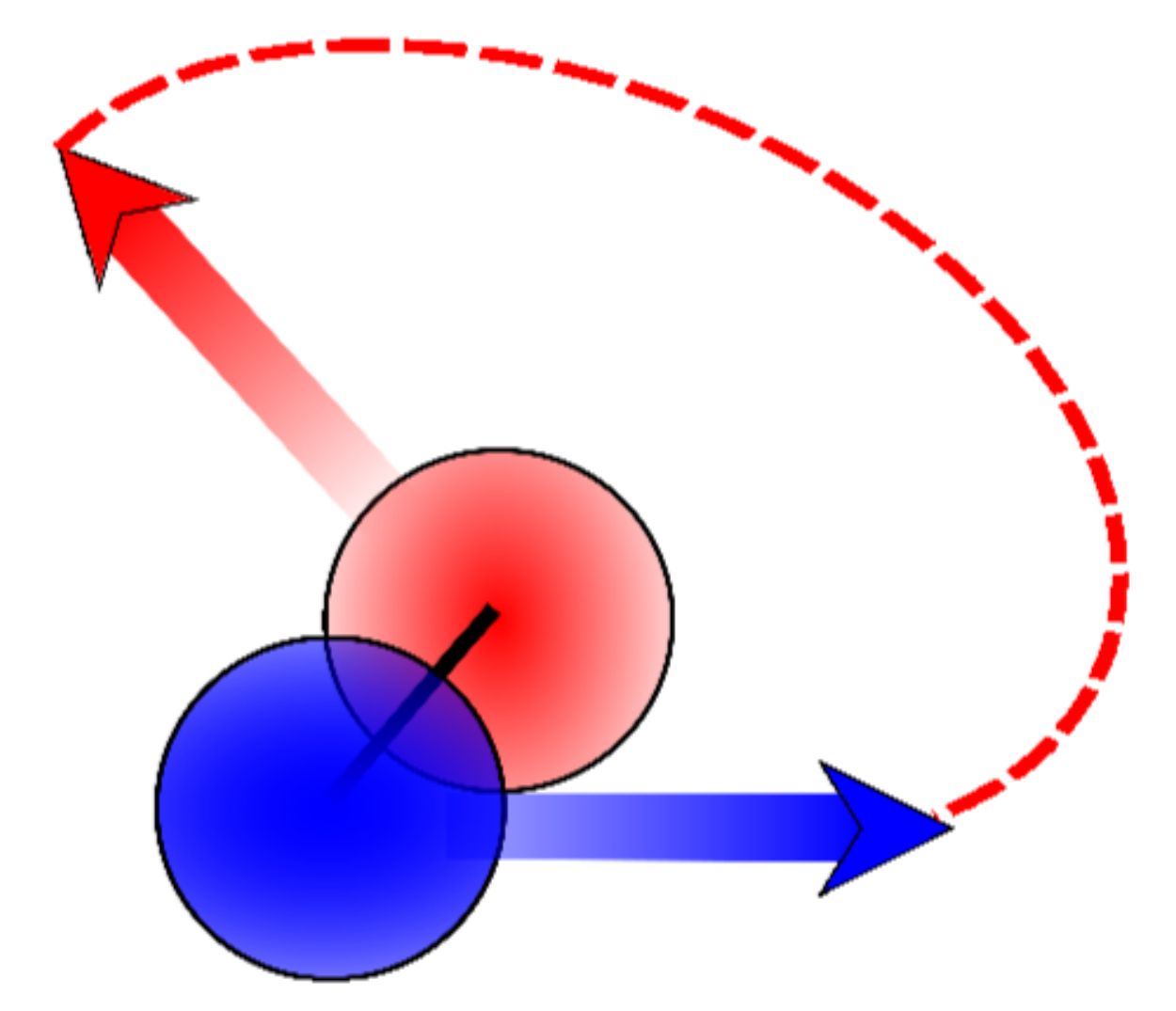}}
    \caption{(Color online) Schematic representation of the $\eta_i$ angle. The 
    sphere represents the C$^\alpha$ atom and the vector points toward the C$^\beta$ 
    postition. The $i^{th}$ residue site is marked with blue and the $(i + 1)^{st}$ is marked with red. }
    \label{fig-5}
  \end{center}
\end{figure}

During the simulations we monitor  the evolution of the  ($\theta_i, \gamma_i$) angles defined
in (\ref{eq:bond}) and (\ref{eq:torsion}), 
and the side-chain C$^\beta$--C$^\beta$ 
torsional angles which  we denote by $\eta_i$ (schematically shown in the Figure \ref{fig-5}).
For this, the C$^\beta$ positions are calculated by converting the coarse-grained UNRES 
trajectories into all-atom trajectories using an optimal dipole aligment algorithm developed in \cite{kazmierkiewicz_2002,kazmierkiewicz_2003}. 

We choose the 10$^{th}$ residue along the 1GAB backbone as a reference point: We have observed 
that it remain highly inert throughout our simulations. In particular, this residue does not seem to be
affected by the kink propagation which we instigate. Starting with this reference site, we 
combine the side-chain C$^\beta$--C$^\beta$  torsional angles $\eta_i$ into 
the following accumulated {\it total} side-chain torsion angle $\hat \eta_i$,  
\begin{equation}
\hat \eta_i= \sum_{k=10}^i \eta_k
\label{eq:etaprime}
\end{equation}
During the simulations, we also monitor two quantitative criteria  to assess whether the  
displaced kink has reached  its native location. 
One of the criteria is the RMSD between the simulated structure and the 
PDB structure 1GAB. Following \cite{krupa_2013}, we demand that for a native state, the
RMSD should be less than 4\AA. The second criteria is based on the ($\theta_i,\gamma_i$) profile
of the kink, as shown in  figures \ref{fig-3}. 
We demand that at the end, the  kink is located within
one lattice site from the native location between Ala$^{35}$-Val$^{38}$. 

Finally, to estimate the height of the Peierls-Nabarro barrier, we evaluate 
the energy of the fragment Gln$^{30}$-Val$^{38}$ using the UNRES energy function.

\section{Results and discussion}

\subsection{Multiple pathways}
\begin{figure}
  \begin{center}
    \resizebox{8cm}{!}{\includegraphics[]{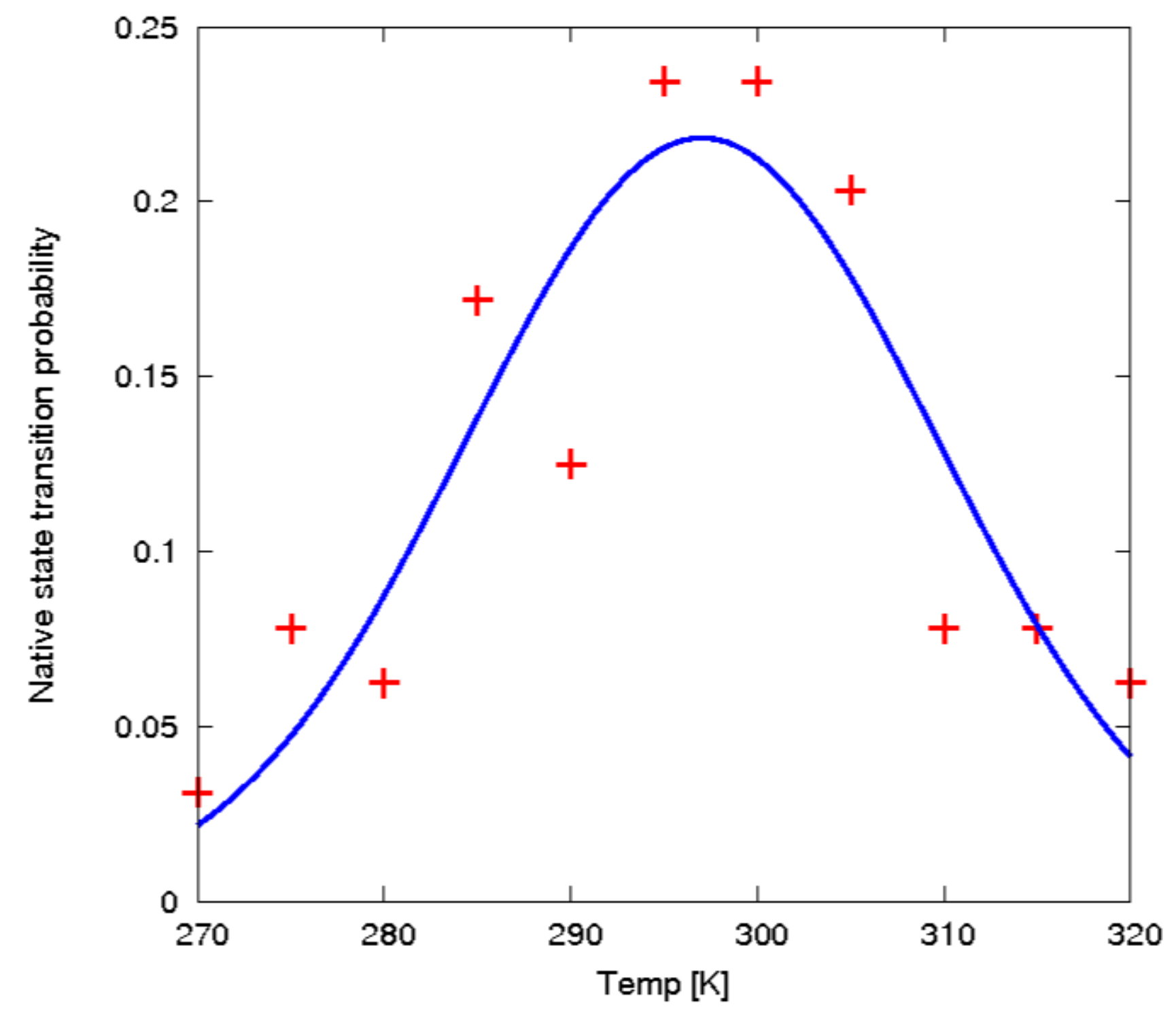}}
    \caption{ (Color online) Fraction of simulations in which kink propagates to the native position, 
    as a function of temperature. The red dots are the measured fraction, the blue line is a gaussian fit.}
    \label{fig-6}
  \end{center}
\end{figure}

According to  \cite{krupa_2013}, when the 1GAB folding is simulated at around 295K using
UNRES force field with replica-exchange molecular dynamics, over 90\% of structures reach the
native state.  However, in our UNRES simulations, using canonical molecular dynamics and
{\it with our initial condition} where the second loop fragment has been parallel translated
by five residues towards the first and then released at rest,  we find that at 295K 
temperature only around 25 per 
cent of structures reach the native state. In figure  \ref{fig-6} we show the
probability for our initial configuration to reach the native state, as a function of temperature. 
The probability has a maximum at around 295K. 
%
%
%
%
%
%
%
%
%
%
We find that at low temperatures, well below 295 K, the Peierls-Nabarro energy barrier becomes too high for 
the kink to leap over it, only by the thermally induced kinetic energy. When the temperature increases and the
amplitude of thermal fluctuations in the kink is larger, the barrier crossing becomes more and more frequent.
But when temperature reaches values which are well above 295 K, the unfolding processes
which are driven by increased thermal fluctuations in the protein lattice structure
start becoming increasingly prominent and eventually the protein unfolds. 

\begin{figure}
  \begin{center}
    \resizebox{8cm}{!}{\includegraphics[]{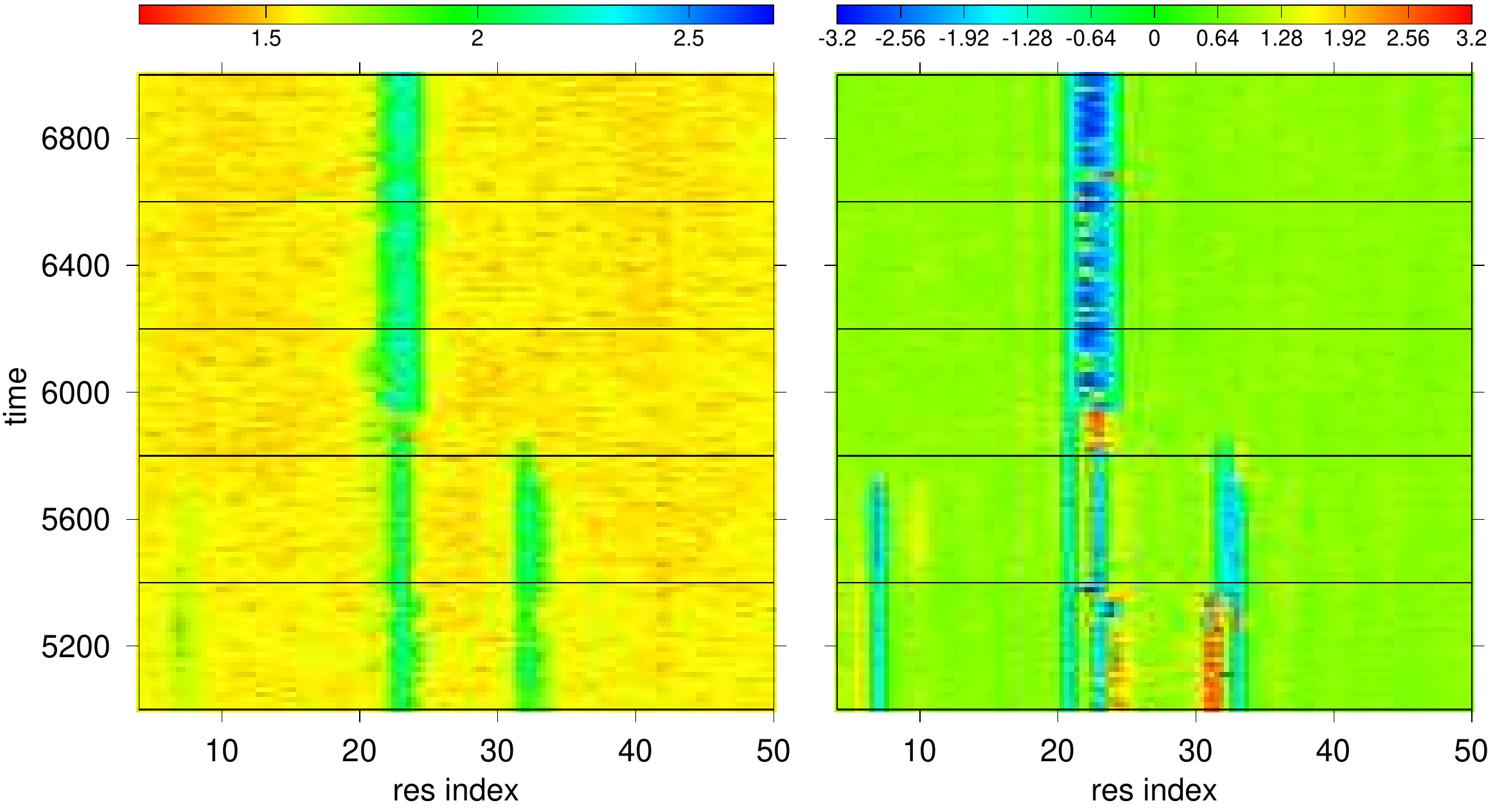}}
    \caption{(Color online) The changes of virtual valence angle $\theta$  (left) and torsional angle $\gamma$ (right), 
    as a function of simulation time and residue index during a process of kink disintegration.   }
    \label{fig-7}
  \end{center}
\end{figure}

\begin{figure}
  \begin{center}
    \resizebox{8cm}{!}{\includegraphics[]{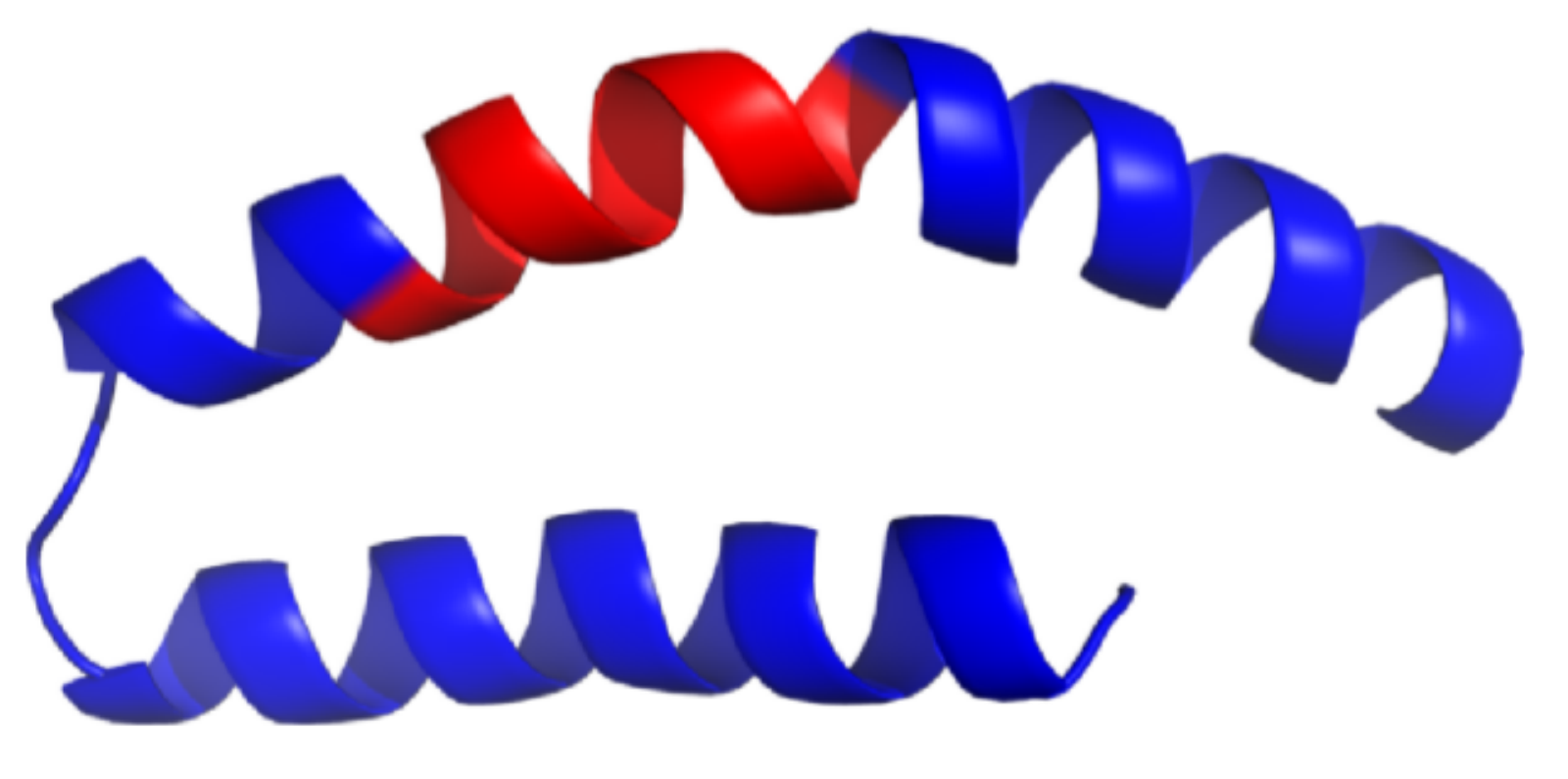}}
    \caption{(Color online) The structure of missfolded 1GAB protein after kink dissolution. The red color denotes the initial position of the parallel translated kink.
   }
    \label{fig-8}
  \end{center}
\end{figure}

\begin{figure}
  \begin{center}
    \resizebox{8cm}{!}{\includegraphics[]{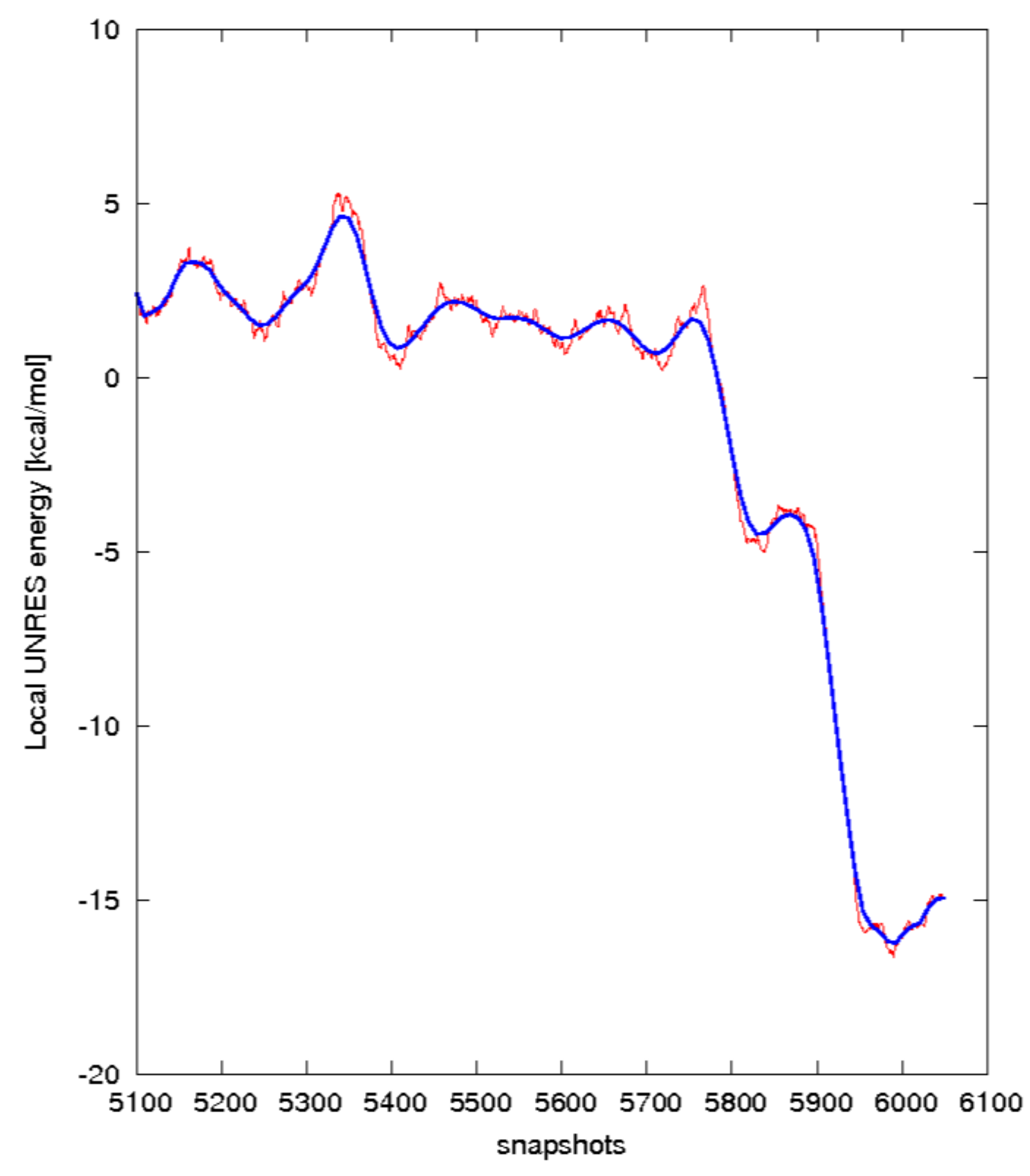}}
    \caption{(Color online) The local energy change during kink dissolution, evaluated over residues 29-40. 
    The red line is the moving average over 50 snapshots and the blue line is a smoothed moving average.   }
    \label{fig-9}
  \end{center}
\end{figure}

We also find that at temperatures well below 295 K, when the Peierls-Nabarro energy barrier becomes too
high for the kink to cross over, the most frequently occurring pathway 
entails a disintegration of the kink; see Figure \ref{fig-7}. 
Along this pathway, the parallel transported loop misfolds into a helical shape, and 
at the end we have a  helix-bend-helix configuration akin the one shown in figure \ref{fig-8}.
We find that the  energy barrier for this transition is quite low, as shown in figure \ref{fig-9}. 

\subsection{Kink propagation}

It should be noted that in our simulations each trajectory 
describes kink propagation in a  somewhat different manner, no two trajectories are fully  identical. 
There are also trajectories with kink-kink interactions. 
Only the general aspects 
of the most commonly occurring pathways are detailed in the sequel.

\begin{figure}
  \begin{center}
    \resizebox{8cm}{!}{\includegraphics[]{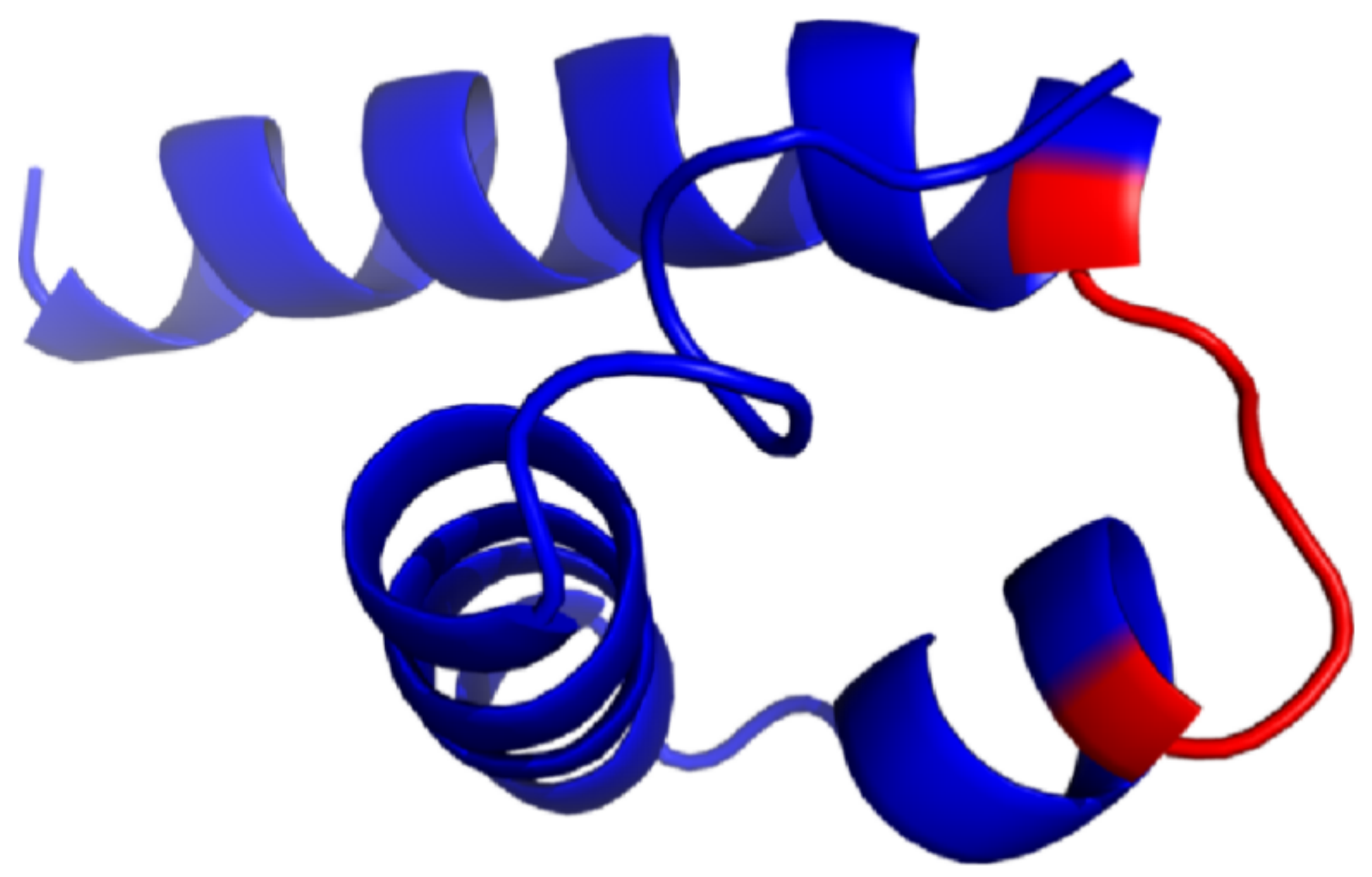}}
    \caption{(Color online) The structure of 1GAB protein after the initial transition with tension reduction. 
    Note that the third helix becomes perpendicular to second helix which reduces the height of the 
    Peierls-Nabarro energy barrier.
   }
    \label{fig-10}
  \end{center}
\end{figure}

The kink propagation commences with a process of tension reduction:  In the initial stage,  
the first and third helix are aligned in parallel, next to each other. 
Interactions between the ensuing side chains are present, 
causing a tension which hinders the 
propagation of the parallel transported kink.
This tension becomes relaxed by  a deformation that orients  the first and third helix
perpendicular to each other (see figure \ref{fig-10}).
This lowers the Peierls-Nabarro barrier, enabling the kink to start propagating. 

\begin{figure}
  \begin{center}
    \resizebox{8cm}{!}{\includegraphics[]{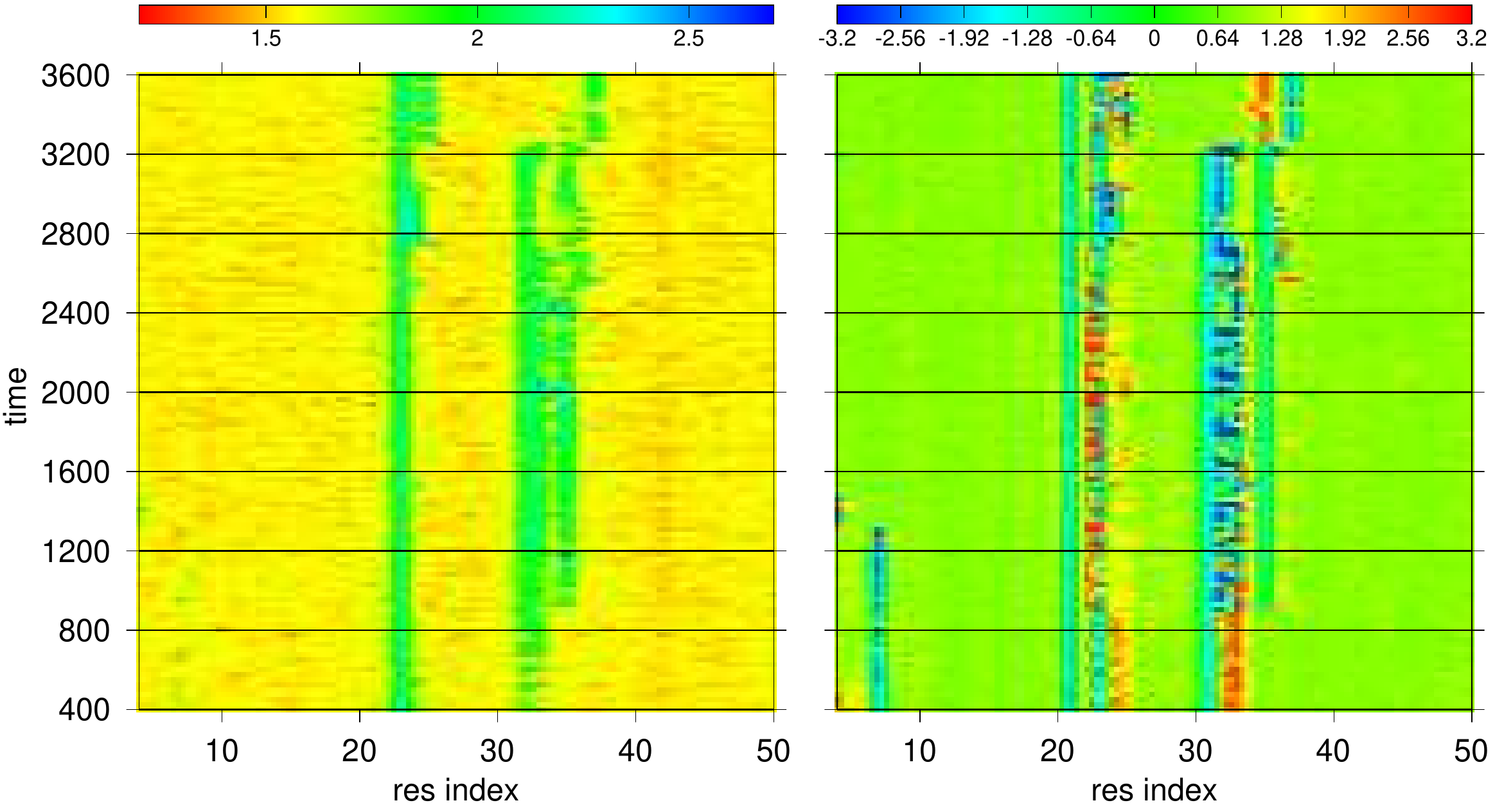}}
    \caption{(Color online) The variations of $\theta$ (left) and $\gamma$ (right), as a function of time and 
    residue index, during the kink propagation towards the native position; the color-coding is defined on top.
       }
    \label{fig-11}
  \end{center}
\end{figure}

\begin{figure}
  \begin{center}
    \resizebox{8cm}{!}{\includegraphics[]{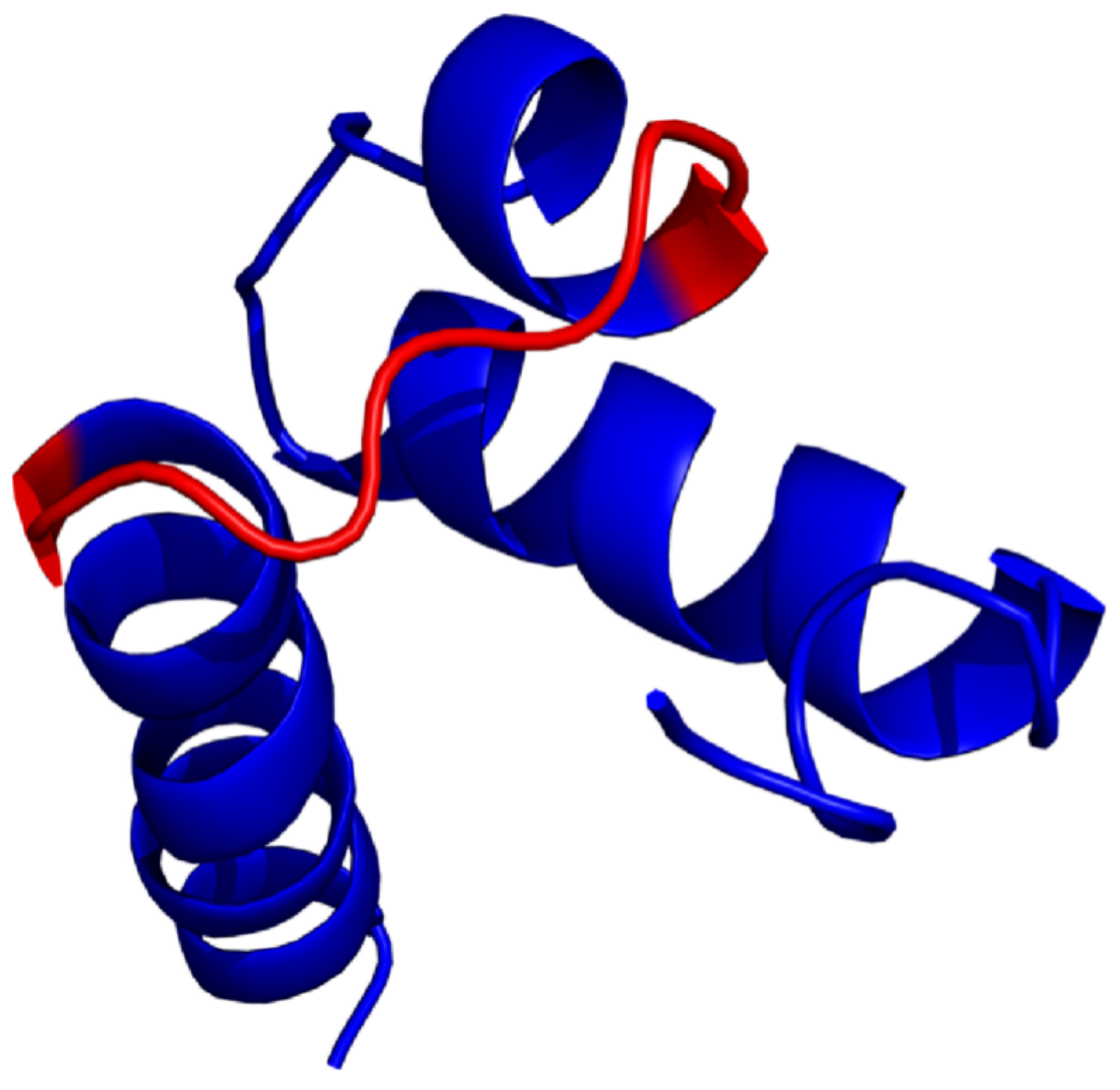}}
    \caption{(Color online) The structure of 1GAB protein when the kink is prolongated.   }
    \label{fig-12}
  \end{center}
\end{figure}

Subsequently, we observe that the kink propagates along the chain 
by {\it reptation}: There is an initial prolongation 
of the kink structure, an example is shown in  figure \ref{fig-11}. 
In this particular example  the kink extends from residue 
30 to residues 36-38, during the UNRES time
period between  $\approx$ 1000 and $\approx$ 3200. The prolongated kink corresponds to an extended 
loop region, which is shown in figure \ref{fig-12}; we also
note that the third helix is perpendicular to both the first and the second helix. 
This enhances the conformational flexibility of the loop fragment which enables an increase in the accumulation of
thermally induced kinetic energy,  boosting the kink to more easily cross over the energy barriers 
towards the final position. In the final stage the third helix then returns to a posture where it is (anti)parallel to the other two helices,
and the kink assumes its native location.

\subsection{Side-chain motion}

\begin{figure}
  \begin{center}
    \resizebox{8cm}{!}{\includegraphics[]{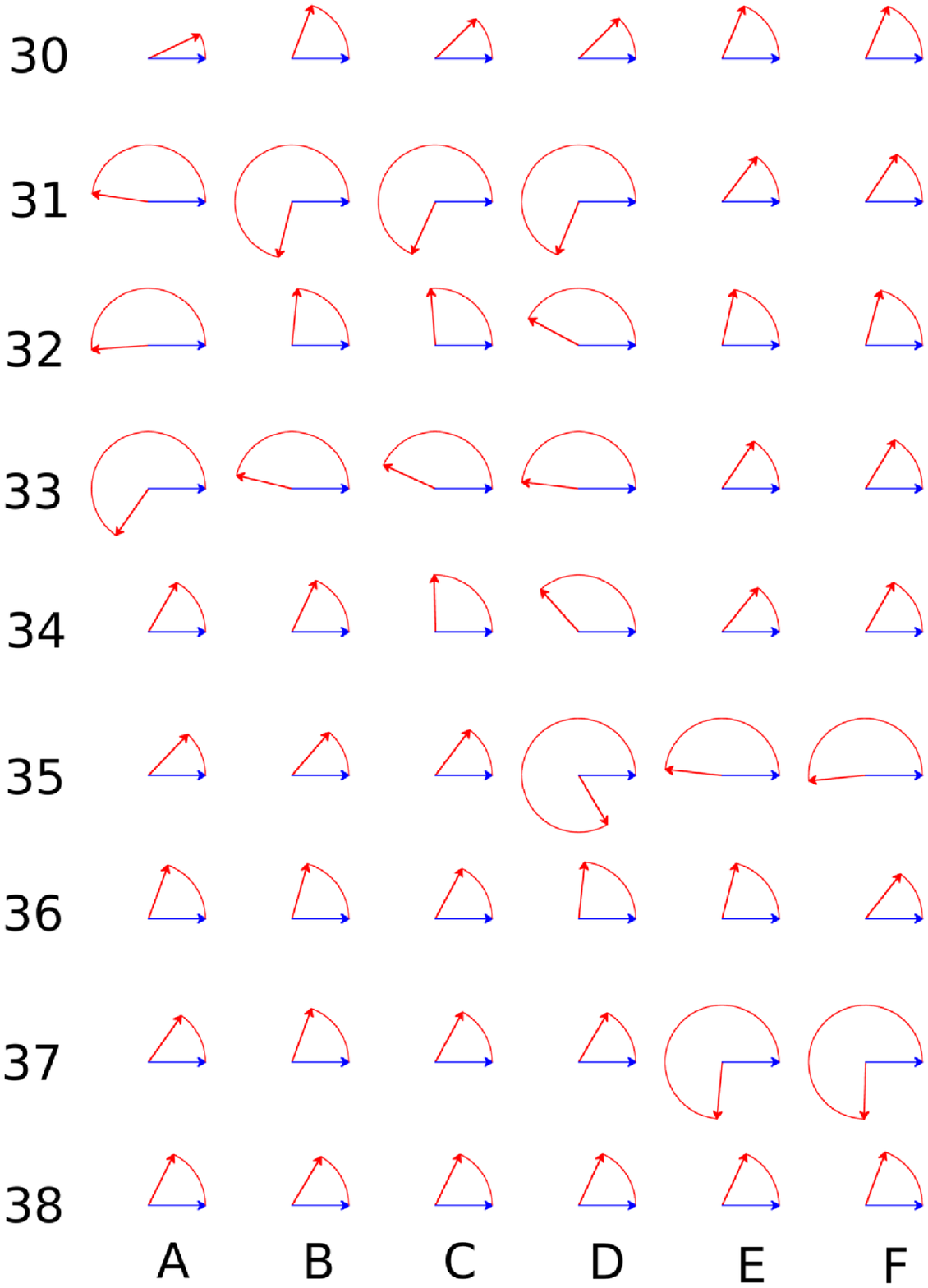}}
    \caption{(Color online) The times series (columns A-F) of $\eta$ angles (residues 30-38).
   }
    \label{fig-13}
  \end{center}
\end{figure}

The kink propagation is strongly correlated with the side-chain motions, as seen in the Frenet frames. 
For this, we first analyze the  C$^\beta$--C$^\beta$  torsional angles $\eta_i$.
The initial values of these angles, for the parallel transported kink,
are shown in column A of figure \ref{fig-13}. 
%
This figure shows that the value of $\eta_i$ along a helical structure 
is $\eta_i  \approx\ $0.8 radians whereas along the kink the $\eta_i$ deviate from this helical value.
In figure \ref{fig-13} column B we observe how  the values of $\eta_i$ 
remain largely intact during the initial phase, that of tension reduction,  
when the relative positions of the helices along the chain do not change.

During the second stage, that of reptation when the kink becomes prolongated, corresponding to the
columns  C and D in figure \ref{fig-13}, the values of the 
$\eta_i$ angle deviate from the helical value $\eta \approx\ $0.8 radians, over an expanded range. 
There is a clear counter-clockwise relative rotation of the side-chains along the chain, 
correlating with the prolongation of the reptating  kink along the chain.

In the final stage, as the third helix assumes its (anti)parallel orientation with the first and second helices,
 the length of the reptating kink  becomes  contracted into a region that extends over the residues 35-37 as  
 the kink moves to its  native position. This can be seen in column E and F of  Figure \ref{fig-13}.

\begin{figure}
  \begin{center}
    \resizebox{8cm}{!}{\includegraphics[]{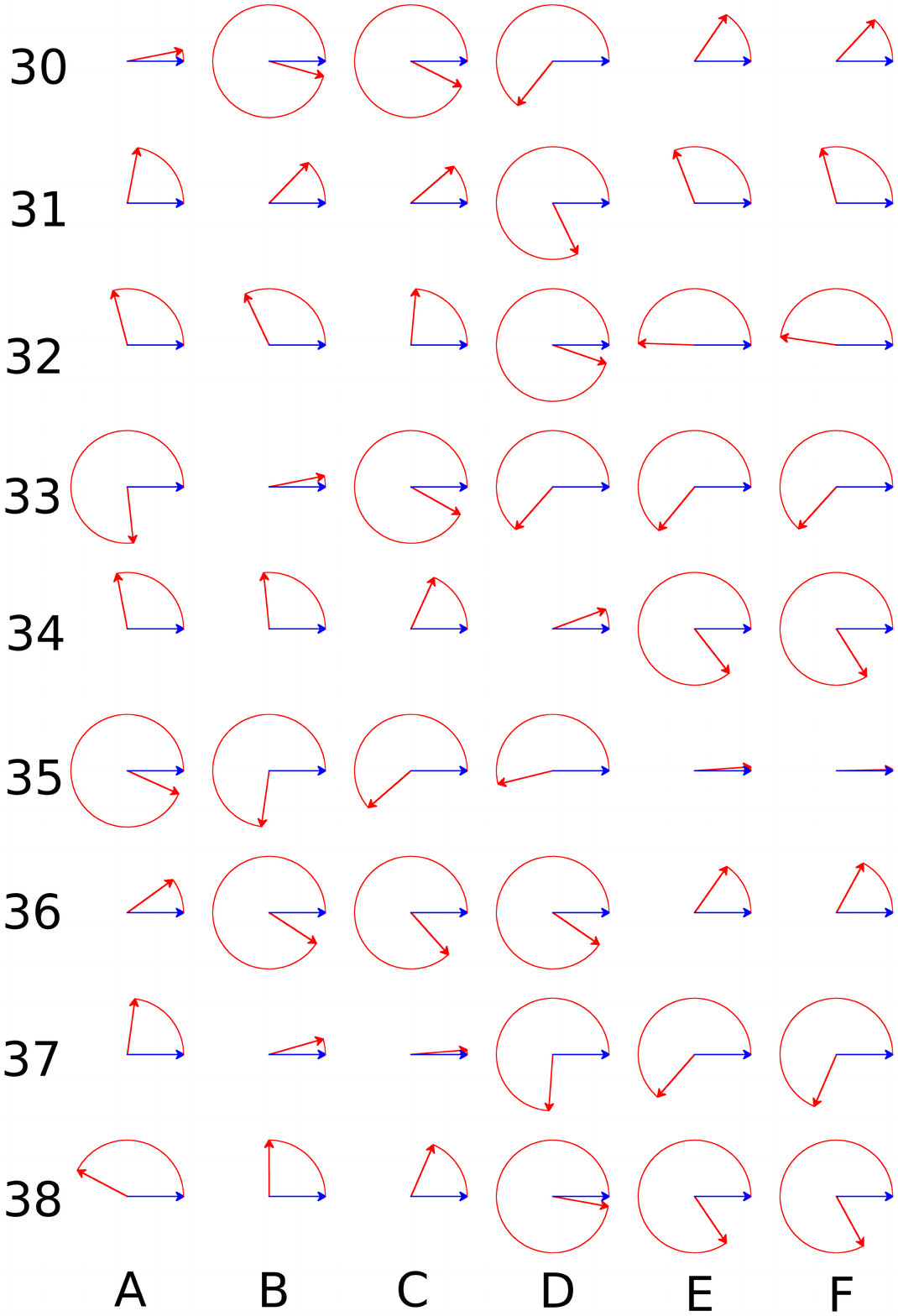}}
    \caption{(Color online) The times series (A-F) of $\hat\eta$ angles (residues 30-38).   }
    \label{fig-14}
  \end{center}
\end{figure}

Additional structure can be identified in terms of the accumulated  total torsion angles $\hat \eta$, that we 
introduced in equation (\ref{eq:etaprime}):  Unlike the C$^\beta$--C$^\beta$  torsional angles $\eta_i$ which
are defined locally, the total angles $\hat \eta_i$ contain both local and global information.
In figure \ref{fig-14} 
%
%
%
we show the values of $\hat\eta_i$ that correspond to the values of $\eta_i$ in figure \ref{fig-13}.

For example,  compare the  first row (residue 30)  in figures \ref{fig-13} and \ref{fig-14}: In figure \ref{fig-13} 
there is no change, but in colums A-C of row 1 in figure \ref{fig-14} we see a substantial 
counter-clockwise rotation. Thus, we conclude that the kink propagation correlates with a global helical counter-clockwise
rotation of the side-chains, that occurs prior to the position of the propagating kink.  

Note that in both figures \ref{fig-13} and \ref{fig-14} there is no relative side-chain rotation
visible between the two last columns. The value of $\eta_{36}$ (angle between reside 36 and 37) remains  essentially
constant throughout our simulations. This orientational locking of the two side-chains with each other
reflects a strong repelling interaction between the two lysines which are located at sites 35 and 37.
This makes the two residues 36  and 37 
to rotate in tandem.

\subsection{Height of the Peierls-Nabarro barrier}
\begin{figure}
  \begin{center}
    \resizebox{8cm}{!}{\includegraphics[]{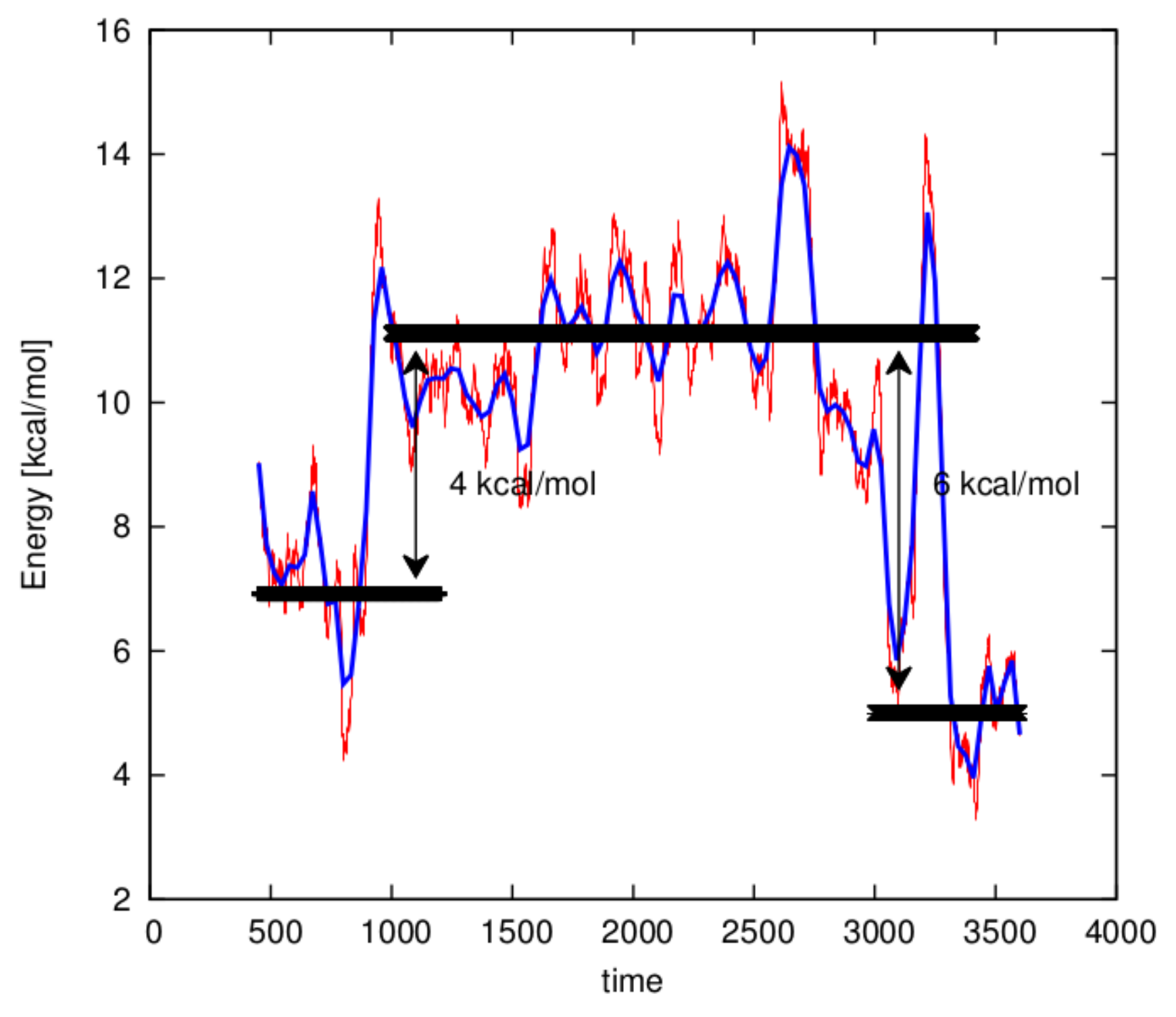}}
    \caption{(Color online) The local energy changes during the kink propagation, evaluated over the residues 29-40. The red line represents the moving average over the 50 snapshots and the blue line is a smoothed moving average.
   }
    \label{fig-15}
  \end{center}
\end{figure}

We have monitored  the change of local  energy  associated with the kink propagation; 
see figure \ref{fig-15}.
We observe that the difference in the energy which is localized at the initial  
parallel transported kink, and the energy which is localized at the final natively placed
kink,  is very small. Our estimate for the energy difference between the two kinks is 
around  1-2 kcal/mol. However, the energy barrier that the parallel transported kink  needs to cross
over in order to reach is native location, is much higher. From figure \ref{fig-15} we estimate that
the height of the Peierls-Nabarro barrier is around 4-6 kcal/mol. This estimate is comparable, albeit sligtly lower than
 the result obtained  in \cite{krokhotin_2014}. It appears that the major contribution towards the height
of the barrier correlates with the elongation of the kink, during  reptation.

\section{Summary}

In this article  we have analysed the propagation of a kink along protein backbone. The process is much more 
complex than a simple translation,  it involves both local and global deformations.  In particular, the  Peierls-Nabarro 
barrier crossing involves a reptation-like movement, with the kink propagating by a combination of prolongations and contractions,
and with a very sudden character. We note that this is in line with the analysis of soliton  
propagation along a cyclic  molecular chain \cite{brizhik_2000}.

We have estimated the height of the ensuing Peierls-Nabarro barrier to be around 4-6 kcal/mol, in the case of the second loop of
1GAB. We observe that this is remarkably close to the  energy gap in ATP to ADP hydrolysis process. 

It appears that the present study is the first to address the relevance of Peiers-Nabarro barrier to protein 
folding. Moreover, we have only considered a particular example. However, most protein loops can be
described in terms of different parametrizations of the kink solution which is supported by (\ref{nlse}), (\ref{tauk}).
Thus we trust that the phenomenon we have revealed is not limited to the specific example, but  relates to a general
mechanism how proteins fold and unfold. 

Finally, we note that for a detailed investigation of loop propagation and 
the Peierls-Nabarro barrier, an all-tom molecular dynamics simulation should be performed 
for example along the lines of 
\cite{cossio_2010}. Such simulations, in particular over extended temperature range as described here 
remain highly time consuming, thus we postpone a detailed all-atom investigation to a future article.

\section{Acknowledgments}

AKS reaserch at Uppsala University was supported by Swedish Institute scholarship,
AJN  acknowledges support from Region Centre 
Rech\-erche d$^{\prime}$Initiative Academique grant, Sino-French 
Cai Yuanpei Exchange Program (Partenariat Hubert Curien), 
Vetenskapsr\aa det, Carl Trygger's Stiftelse f\"or vetenskaplig forskning, 
and  Qian Ren Grant at BIT. This research has been supported by
an allocation of advanced computing
resources provided by the National Science Foundation
(http://www.nics.tennessee.edu/),
and by the National Science Foundation through
TeraGrid resources provided by the Pittsburgh Supercomputing Center.
Computational resources were also provided by
(a)
the supercomputer resources at the Informatics Center of the Metropolitan Academic Network (IC MAN)
in Gda\'nsk,
(b)
the 624-processor Beowulf cluster at the Baker Laboratory of
Chemistry, Cornell University, and
(c)
our 184-processor Beowulf
cluster at the Faculty of Chemistry, University of Gda\'nsk.

\end{document}